\documentclass[aps,prd,reprint,superscriptaddress]{revtex4-2}

\usepackage{comment}
\usepackage[utf8]{inputenc} 
\usepackage{graphicx,color,overpic,mathtools}
\usepackage{amsthm,amsmath,amssymb,hyperref,mathrsfs}
\usepackage{braket,bm,bbm,setspace}
\PassOptionsToPackage{normalem}{ulem}
\usepackage{ulem} 
\usepackage{physics}
\usepackage{float}
\usepackage[makeroom]{cancel}
\usepackage[english]{babel}
\usepackage{graphicx}
\usepackage{soul}
\graphicspath{ {images/} }
\addto\captionsspanish{}
\hypersetup{
    colorlinks=false,
    pdfborder={0 0 0},
} 

\begin{document}

\title{Inversion of statistics and thermalization in the Unruh effect}

\author{Julio Arrechea}
\email{arrechea@iaa.es}
\affiliation{Instituto de Astrof\'isica de Andaluc\'ia (IAA-CSIC),
Glorieta de la Astronom\'ia, 18008 Granada, Spain}

\author{Carlos Barcel\'o}
\email{carlos@iaa.es}
\affiliation{Instituto de Astrof\'isica de Andaluc\'ia (IAA-CSIC),
Glorieta de la Astronom\'ia, 18008 Granada, Spain}

\author{Luis J. Garay}
\email{luisj.garay@ucm.es}
\affiliation{Departamento de F\'{\i}sica Te\'orica and IPARCOS, Universidad Complutense de Madrid, 28040 Madrid, Spain}  
\affiliation{Instituto de Estructura de la Materia (IEM-CSIC), Serrano 121, 28006 Madrid, Spain}

\author{Gerardo García-Moreno}
\email{gerargar@ucm.es}
\affiliation{Departamento de F\'{\i}sica Te\'orica and IPARCOS, Universidad Complutense de Madrid, 28040 Madrid, Spain}
\affiliation{Instituto de Astrof\'isica de Andaluc\'ia (IAA-CSIC),
Glorieta de la Astronom\'ia, 18008 Granada, Spain}

\begin{abstract}
{
We derive a master equation for the reduced density matrix of a uniformly accelerating quantum detector in arbitrary dimensions, generically coupled to a field initially in its vacuum state, and analyze its late time regime. We find that such density matrix asymptotically reaches a Gibbs state. The particularities of its evolution towards this state are encoded in the response function, which depends on the dimension, the properties of the fields, and the specific coupling to them. We also compare this situation with the thermalization of a static detector immersed in a thermal field state, pinpointing the differences between both scenarios. In particular, we analyze the role of the response function and its effect on the evolution of the detector towards equilibrium. Furthermore, we explore the consequences of the well-known statistics inversion of the response function of an Unruh-DeWitt detector linearly coupled to a free scalar field in odd spacetime dimensions. This allows us to specify in which sense accelerated detectors in Minkowski vacuum behave as static detectors in a thermal bath and in which sense they do not. 
}
\end{abstract}

\maketitle

\section{Introduction}

It is often stated that the Unruh effect consists in the perception of the Minkowski vacuum as a thermal bath by an accelerated observer \cite{Unruh1976}. This conclusion can be reached through various approaches. Foremost, the comparison between Fock quantizations with respect to Rindler and Minkowski modes for free theories exemplifies this. Although unitary inequivalent, it can be seen that, in arbitrary dimensions, the Minkowski vacuum is related with a state in the Rindler quantization by a Bogoliubov transformation whose coefficients reveal a Planckian population, with thermality appearing because of the tracing over the unobservable Rindler wedge. The Rindler quantization has a free parameter, describing the acceleration of the observer whose proper coordinates are the ones of the Rindler spacetime. The temperature of this thermal population is therefore proportional to the acceleration of this privileged observer \cite{Fulling1972,Davies1974,Unruh1976}. Another formal derivation of this result, which actually holds for arbitrary interacting theories described by a Lagrangian, is via path integral methods. Albeit this derivation assumes a factorization of the Hilbert space which, strictly speaking, cannot be performed \cite{Witten2018}, it succeeds in proving the following result: Expectation values in the Minkowski vacuum of observables supported on one of the Rindler wedges with respect to the Minkowski Hamiltonian can equally be computed as thermal expectation values with respect to the Rindler Hamiltonian. This result is often referred to as thermalization theorem \cite{Lee1986,Friedberg1986}. Furthermore,  for every spacetime possessing a bifurcate Killing horizon, there exists a generalization of the Unruh effect~\cite{Fulling1987,Kay1991,Moretti2010}. From a more axiomatic approach to quantum field theory, the Unruh effect is a direct consequence of the Bisognano-Wichmann theorem \cite{Bisognano1975}. For a recent review on the Unruh effect and some of its applications see, for instance, \cite{Crispino2007}.

Beyond formal analyses relating Minkowski and Rindler quantizations, in an effort to endow the observer with a more definite physical meaning, models of detectors have also been studied in the literature. The simplest one is the Unruh-DeWitt detector model~\cite{Unruh1976} consisting of a pointlike two-level system that responds to vacuum fluctuations of fields. The detector is then allowed to follow an accelerated trajectory, its interaction being turned on and off by a suitable switching function. This type of calculation leads to the definition of the detector {\em response function} (proportional to its probability of excitation per unit of time), which depends only on the trajectory of the detector, the state of the field and the coupling between the field and the detector. Consider for instance a free massless real scalar field linearly coupled to the monopole of the detector. In 4 spacetime dimensions, the response function for a uniformly accelerated detector in the Minkowski vacuum exhibits a Planckian shape: the same that it would exhibit when it follows a static trajectory with the field in a thermal state. A puzzle appears, however, when considering dimensions different than 4: the response function of the accelerated detector in vacuum strongly depends on the number of spacetime dimensions. Throughout this work we will restrict ourselves to study cases of dimension greater than 2 since, for the massless field, there are infrared divergences that require a special treatment. Dimension 3 is special for this model of detector since the Planckian distribution is exactly replaced by a Fermi-Dirac one. Considering dimensions greater than 4 we obtain, for even dimensions, a Planckian distribution modulated by dimension-dependent polynomial factors in the frequency while, for odd dimensions, we obtain a Fermi-Dirac one modulated by dimension-dependent polynomial factors.

The situation is worsened when we consider couplings to an arbitrary integer power of the field. Particularly, for detectors coupled to a free massless real scalar field $\phi$ through an even power of the field, the response function has, for every dimension, a Planckian form modulated by dimension and coupling dependent polynomial factors in the frequency. If the coupling to the detector is done through an odd power of the field, the response function switches between the Planckian and Fermi-Dirac functional forms depending on whether we are in even or odd dimensions, respectively; appearing in both cases modulated by a dimension and coupling dependent polynomial factor in the frequency again.

The phenomenon of statistics inversion was originally noticed by Takagi \cite{Takagi1984,Takagi1985} and it was presented as an obstruction to the thermal interpretation of the Unruh effect. To further complicate the matter, the response function for massive fields was also observed to deviate from a Planckian-like form. For a review and further generalizations where similar phenomena are observed, like an analogous result for fermion fields, see \cite{Takagi1986}. Later pinpointed by Unruh \cite{Unruh1986} without much ado, he claimed that there was no contradiction between the standard interpretation and Takagi's results: the relevant feature was the Boltzmann factor relating the populations of each of the field modes.

In a subsequent work, Ooguri \cite{Ooguri1985} identified the formal reasons for the dependence of this phenomenon on the spacetime dimension. Owing to the Kubo-Martin-Schwinger (KMS) condition~\cite{Kubo1957,Martin1959}, the response function is fixed to be a Planck distribution multiplied by the Fourier transform of the pullback of the commutator to the trajectory of the field~\cite{Ooguri1985}. If the propagator has support only on the light cone, as the Huygens principle guarantees in even spacetime dimensions, the Fourier transform of the commutator necessarily has to be a polynomial. In odd dimensions, the commutator has support also inside the light cone, i.e., there is propagation of signals at speeds lower than the speed of light. This leads to nonpolynomial functions of the frequency (i.e., quotients of exponentials) that compensate the Planckian factor and transform it into a Fermi-Dirac factor. In consequence, the response function is non-Planckian-like in odd dimensions. Notice that these features have nothing to do with the quantum nature of the Unruh effect: the commutator is completely determined by the classical dynamical evolution for the field. 

More recent works \cite{Sriramkumar2002,Sriramkumar2016} allude to the statistics inversion being a mere feature of the detector but not clarifying the concerns put forward by Takagi. The additional polynomial dimension-dependent factors that enter the response function have been argued to represent the effective density of states seen by an accelerated observer~\cite{Ottewill1987}.

It is our aim to shed light into this subject and to fill the gap in comprehension that appears to still be present in the literature. We will argue that the thermal nature of the Unruh effect is, indeed, correct. The commented particularities of the response function have nothing to do with the thermal nature of the Unruh effect. As we will show, this thermal nature is encoded within the Boltzmann factor relating positive and negative frequency response functions, as it is implied by the KMS condition, which all the relevant Wightman functions obey. However, we will also argue that the thermal nature of the Unruh effect does not mean that the same detector accelerating in vacuum and immersed in a static thermal bath will behave exactly in the same way: This is only true in four dimensional spacetime and for detectors linearly coupled to massless free fields, any change in these characteristics leading to huge room for variability in the response function. The Unruh effect is a thermal effect in the sense that a physical detector subjected to uniform acceleration reaches asymptotically a thermal (Gibbs) state, and this is independent from having a strictly Planckian response function. We will derive this universal property by treating a standard Unruh-DeWitt detector as an open quantum system and obtaining its asymptotic time evolution at first order in perturbation theory for times much longer than the thermalization time. As long as the pullback of the Wightman functions obeys the KMS condition, which is the case for the uniformly accelerated trajectory, we will show that the detector thermalizes. Furthermore, we find out that the response function encodes the information about the path that the detector follows within its space of states to reach thermal equilibrium. Although for simplicity we will work with a generic real scalar field coupled to a monopolar detector, the method and results presented in this paper are straightforwardly extendable to other fields, like the fermionic fields considered by Takagi \cite{Takagi1986}, where similar arguments apply.

This paper is structured as follows: Section \ref{Sec:UdW} summarizes computations and results scattered through the literature related to this apparent inversion of statistics. Additionally, we emphasize some generalities related to thermal baths relevant to understand the puzzle that this phenomenon has raised. In Sec.~\ref{Sec:Open} we obtain the asymptotic evolution for the density matrix of the detector and analyze the resulting master equation. Section \ref{Sec:Response} contains the analysis of some specific models and we highlight the differences between the Unruh effect and a static thermal bath. Finally, we summarize our conclusions in Sec.~\ref{Sec:Conclusions}. Throughout the work we use units $\hbar=c=k_{\rm B}=1$ and consider the signature $\textrm{diag} \left(-,+,...,+ \right)$ for the Minkowski metric.

\section{Unruh-deWitt detectors: Thermal vs Planck vs Unruh.}
\label{Sec:UdW}

Our departure point is an arbitrary real scalar field theory, i.e., a field theory obeying Wightman axioms \cite{Haag1992}. This means that we can consider either a massive or massless, interacting or noninteracting field, as long as it defines a quantum field theory in the sense of Wightman. Under such conditions, reconstruction theorems ensure that all the properties of the theory are encoded in the $n$-point functions, commonly referred to as Wightman functions. Thus, we will assume that the whole theory is defined in terms of its Wightman functions, which we take as input data. Furthermore, we model the Unruh-DeWitt detector as a pointlike two-level system whose free Hamiltonian is given by $H_{\textrm{d}} = \omega \sigma_z /2$, with $\sigma_z$ being the diagonal Pauli matrix and $\omega$ the energy gap between the ground and excited states of the detector, which we represent as $ \{ \ket{\text{g}}, \ket{\text{e}} \}$, respectively. We will consider an interaction Hamiltonian between the two-level system and the field $\phi(x)$ which, written in the interaction picture and parametrized by the proper time of the detector $\tau$, reads 
\begin{equation}\label{Eq:IntHam}
    H_{\textrm{I}}(\tau) = \lambda \chi(\tau/\Delta) m(\tau) \otimes \phi^n [ x(\tau)]. 
\end{equation}
Here, $\lambda$ is the coupling constant in terms of which we perform a perturbative expansion; $\chi(\tau/\Delta)$ is the switching function: a square integrable, smooth function $\left(\chi (y) \in C^{\infty} (\mathbb{R})\right)$ controlling the on-and-off switching of the interaction, as well as its duration ($\Delta$ is the width of the interaction window); $m(\tau)$ is the monopole moment of the detector which, written in terms of the $SU(2)$ ladder operators, reads $m(\tau) = e^{i \omega \tau} \sigma_{+} + e^{-i \omega \tau} \sigma_{-}$. For powers $n>1$, physical observables (like the vacuum excitation probability) diverge and need to be tamed by introducing a suitable renormalization scheme. This is a generic feature of interaction Hamiltonians containing arbitrary polynomials of fields. A rigorous treatment~\cite{Hummer2015} of these divergences for $n=2$ reveals that they are analogous to the tadpole diagrams from quantum electrodynamics. Such divergences can be suitably eliminated by applying the normal order prescription to the $\phi^2$ operator appearing in the Hamiltonian. Extending the same prescription to arbitrary $n$ involves the replacement of $\phi^n[x(\tau)]$ by its normal-ordered counterpart in~\eqref{Eq:IntHam}. 

Our aim in this work is to study the thermalization of the detector in different situations. Since thermalization requires that the detector settles with the myriad of interactions in its surroundings, we will need the window of time during which the detector interacts with the field to be greater than any other timescale involved in the problem under consideration (for instance, the thermalization time, which is roughly the inverse of the temperature). For this purpose, it is convenient to consider the so-called infinite adiabatic limit in which the window of time for which the interaction is switched on becomes infinite. This is done by choosing  $\chi(\tau/\Delta)$ within an appropriate family of square integrable functions (such as Gaussian functions) so that, in the limit $\Delta \rightarrow \infty$, it becomes a constant function. This procedure is compulsory in order to manipulate well-defined objects in our construction: the field operators by themselves are distribution-valued operators. We need to smear them with suitable smooth functions to build a meaningful operator acting on the Hilbert space of states. Thus, for all practical purposes, we can set the function $\chi(s) = 1$ in our computations. 

Let us particularize, for a scalar field in $D$ spacetime dimensions, to stationary states and trajectories. Then, the pullback of the $2n$-point  Wightman functions to the trajectory of the detector depends only on the differences between proper times of the detector along its trajectory, and it will hence be denoted by $ \mathcal{W}^{(2n)}_{D} (\tau - \tau')$. The detector probability of excitation per unit time computed to the lowest nontrivial order in perturbation theory turns out to be~\cite{Takagi1986}
\begin{equation}
    R_{+}(\omega) = \lambda^{2} \left|\matrixel{\text{e}}{m(0)}{\text{g}}\right|^2 \mathcal{F}^{(n)}_{D} (\omega).
\end{equation}
Here, the function $\mathcal{F}^{(n)}_{D} (\omega)$ is the response function defined as the Fourier transform of the Wightman function
\begin{equation}
    \mathcal{F}^{(n)}_{D} (\omega) = \int ^{\infty}_{-\infty} d u \mathcal{W}^{(2n)}_{D} (u)e^{-i \omega u}.
\end{equation}
Notice that the response function only depends on properties of the field, on its pullback to the trajectory of the detector, and on the power $n$ of the coupling.
As mentioned before, these response functions for free fields in more than 4 dimensions exhibit departures from Planckianity (details are in Sec.~\ref{Sec:Response}) and can even show statistic inversion \cite{Takagi1986}.  

In this context, apart from the transition to the excited state $(\ket{\text{g}} \rightarrow \ket{\text{e}})$ one needs considering the decay rate $(\ket{\text{e}} \rightarrow \ket{\text{g}})$ per unit time \cite{Takagi1986,Unruh1986}. The probability of decay per unit time is given by
\begin{equation}
    R_{-}(\omega) = \lambda^{2}\abs{\matrixel{\text{g}}{m(0)}{\text{e}}}^2 \mathcal{F}^{(n)}_{D} (- \omega).
\end{equation}
The expressions $R_{+}$ and $R_{-}$ derived for uniform acceleration are relevant also for nonuniform accelerations, whenever the change in acceleration is adiabatic.

The two stationary states and trajectories we will consider are: a static detector immersed in a thermal state for the field, i.e., a state whose Wightman functions obey the KMS condition with respect to an inertial time parameter, and an uniformly accelerated detector in Minkowski spacetime, where the Wightman functions obey the KMS condition with respect to the Rindler time.  The detector thermalizes if the quotient between the populations of the excited and ground states of its reduced density matrix is related by a Boltzmann factor. For a two-level system, it is straightforward to see that the quotient between populations necessarily agrees with the quotient between excitation and decay ratios, since only two states are available. However, let us stress that proving that this quotient between ratios is related by a Boltzmann factor does not correspond to a direct computation of the quotient between populations. In Sec.~\ref{Sec:Open} we will derive this relation explicitly. For the moment, consider the quotient between excitation and decay ratios
\begin{equation}
\label{boltz}
    \frac{R_{+} (\omega)}{R_{-} (\omega)} = \frac{\mathcal{F}^{(n)}_{D} ( \omega)}{\mathcal{F}^{(n)}_{D} (- \omega)}= e^{- \beta \omega},
\end{equation}
where $\beta$ is the inverse of the KMS temperature $T$. Relation \eqref{boltz} is guaranteed for the trajectories under consideration since it is implied by the Fourier transform of the KMS condition \cite{Fewster2016}
\begin{equation}
    \mathcal{W}^{(2n)}_{D} (u- i \beta) = \mathcal{W}^{(2n)}_{D} (-u).
\end{equation}
Thus, we refer to the second equality in Eq. \eqref{boltz} as the Fourier-transformed KMS condition. 

The specific profile (Planckian-like or not) of the response function is not directly related to the thermalization of the detector at sufficiently long times [notice that the number of response functions that obey Eq. \eqref{boltz} is actually infinite]. A Planckian response is characteristic of free field theories in a thermal state. Turning on self-interactions of the field makes the response function non-Planckian. This is a general feature of thermal baths and, as such, is independent of the Unruh effect. Indeed, if we devise a system comprised of a detector immersed in the thermal bath of a field with involved interactions, the response function will show a substantial dependence on the couplings to the field. Nonetheless, the asymptotic quotient of ratios will still be of the form (\ref{boltz}). It is the information about the coupling between the field and the detector the one that becomes reflected in the particular form of the response function, and disappears from the quotient \eqref{boltz}. Thus, the standard interpretation of the Unruh effect as a thermal bath is not contradictory with the particularities of the response functions found by Takagi \cite{Takagi1986}. 

In the next section we will explore the effect of the response function on determining the path that the detector follows towards thermal equilibrium. This discussion requires additional information besides the ratios here exposed. This additional information is precisely the time evolution of the earlier mentioned populations of the reduced density matrix for the detector.

\section{Unruh-deWitt detector as an open system}
\label{Sec:Open}
In this section we will treat the Unruh-DeWitt detector as an open quantum system in which the quantum field plays the role of the environment. We will obtain and study an asymptotic expansion for the density matrix of the detector under the interaction Hamiltonian \eqref{Eq:IntHam}. The components of this density matrix $\rho$ are $ \left\{ \rho_{\text{gg}}, \rho_{\text{ee}}, \rho_{\text{ge}} = \rho_{\text{eg}}^{*}\right\} $, corresponding to the populations of the ground and excited states and the quantum coherences, respectively. 

Let us advance some aspects concerning the approximations involved in deriving the evolution equation for the density matrix of the detector subsystem. The equation will be obtained under the Born-Markov approximation \cite{Rivas2012}, which involves two approximations intertwined to some extent. First, we focus on the regime of weak coupling, which requires $\lambda$ to be small. This weak coupling guarantees that our perturbative expansion captures the relevant effects coming from the interaction between both subsystems. 
Furthermore, it is also crucial in ensuring that the evolution of the detector is Markovian, i.e., that its reduced density matrix can be described by a differential equation that is local in time and its evolution operators constitute a dynamical semigroup \cite{Rivas2012}. This second aspect corresponds to the so-called Markov approximation and it is achieved at the expense of obtaining a ``coarse-grained" reduced density matrix for the detector~\cite{Breuer2002}. This coarse graining implies that the solutions to the master equation will be unable to capture fast oscillations of the density matrix at short timescales compared to the typical timescale of correlations of the field theory under consideration. Furthermore, we recall that the regime which we are working and the approximation that we are making exclude the possibility of analyzing transient effects like the anti-Unruh effect \cite{Brenna2015,Garay2016}.

Now, we proceed to derive the master equation describing the evolution of the reduced density matrix $\rho$ for the detector subsystem. We work out the first nontrivial order in perturbation theory, that is, $\order{\lambda^2}$. The field state acquires corrections at order $\order{\lambda^2}$, and thus, these modifications influence the density matrix of the detector at higher orders in perturbation theory. Thereby, for the purpose of analyzing $\order{\lambda^2}$ corrections to the reduced density matrix of the detector, it is safe to consider that the field remains in its vacuum state. By itself, it is usually called Born approximation
\begin{equation}\label{born}
  \rho_{\textrm{tot}} (\tau) = \rho (\tau) \otimes \rho_{\phi} (0) + \order{\lambda^4}.
\end{equation}
Under the Born approximation, the evolution of the density matrix is described by the following integro-differential equation in the interaction picture \cite{Moustos2016,Benatti2004}:
\begin{equation}\label{Eq:rho}
\begin{split}
    \Dot{\rho} =  \lambda^2 \int_{0}^{\tau} d \tau' \left[ m(\tau') \rho(\tau') m(\tau) - m(\tau) m(\tau') \rho (\tau') \right]\\
    \times  \mathcal{W}^{(2n)}_{D} (\tau - \tau')\\
    + \lambda^2 \int_{0}^{\tau} d \tau' \left[ m(\tau) \rho (\tau') m (\tau') - \rho( \tau') m(\tau') m(\tau) \right]\\
    \times \mathcal{W}^{(2n)}_{D} (\tau' - \tau) + \order{\lambda^4}.
\end{split}
\end{equation}
The $\rho_{\text{gg}}$ and the $\rho_{\text{ee}}$ components of the density matrix are coupled through the equations of motion, but the off-diagonal component is decoupled from diagonal elements and obeys a single closed differential equation. For the purpose of analyzing the populations of the detector, it suffices to consider the evolution of the diagonal components 
\begin{equation}\label{Eq:rhodiag}
\begin{split}
    \Dot{\rho}_{\text{ee}} (\tau) =  - \lambda^2 \int_{0}^{\tau} d \tau' \left[ A^{(n)}_{D}(\tau - \tau')  \rho_{\text{ee}} (\tau') \right.\\
    \left. - B^{(n)}_{D}(\tau - \tau')  \rho_{\text{gg}}(\tau') \right], \\
    \Dot{\rho}_{\text{gg}} (\tau) =  -\lambda^2 \int_{0}^{\tau} d \tau' \left[ B^{(n)}_{D}(\tau - \tau')  \rho_{\text{gg}}(\tau') \right. \\
    \left. -  A^{(n)}_{D}(\tau - \tau')  \rho_{\text{ee}} (\tau') \right],
\end{split}
\end{equation}
with the functions $A^{(n)}_{D} (s), B^{(n)}_{D} (s)$ defined as 
\begin{equation}
\begin{split}
    & A^{(n)}_{D} (s) = 2 \textrm{Re} \left(  \mathcal{W}^{(2n)}_{D} (s) e^{i \omega s} \right), \\
    & B^{(n)}_{D} (s) = 2 \textrm{Re} \left( \mathcal{W}^{(2n)}_{D} (s) e^{-i \omega s} \right),
\end{split}
\end{equation}
where $\textrm{Re}$ denotes the real part of the function.  

We can now make one last approximation relying on the fact that the Wightman function is highly peaked around $\tau - \tau' = 0$ (for $D\geq3$) and that it decays fast to zero, being $\beta$ (the KMS parameter, i.e., the inverse of the temperature) the characteristic size of the region where the function is non-negligible.

Although it is a well-known fact that the pullback of the thermal Wightman function to a static trajectory shows an exponential fall off with decay constant $\sim \beta^{-1}$, it is not clear whether this property extends to accelerated trajectories in vacuum without further assumptions. It is convenient to pause at this point to discuss why the pullback of the Wightman function obeys such a fast decay property for the accelerated trajectory in vacuum. Notice that, for a relativistic quantum field theory in flat space, Poincaré invariance ensures that the two-point Wightman function depends on the Minkowski spacetime interval between the two points $W^{(2)}_{D} (\abs{\Delta x})$. Let us now assume that we have a Wightman function that decays at least polynomially as $\abs{\Delta x} \rightarrow \infty$, which is a fairly reasonable assumption. For spacelike separation, this is just the clustering property typically asked for the Wightman functions to define a meaningful theory. For timelike separation it is an additional assumption, but it is obeyed by all of the Wightman functions we know of, for instance the ones corresponding to a free theory. Actually, particularizing to the pullback of the Wightman function to a trajectory with acceleration $a$, for which $\abs{\Delta x}^2  = 4 \sinh^2 \left( a \Delta \tau /2 \right) /a^2$, the polynomial fall off implies that the Wightman function is bounded by an exponential
\begin{equation}\label{falloff}
    \abs{\mathcal{W}_{D+1}^{(2)} (u)} \leq K  \exp \left( - A u \right),
\end{equation}
with $A, K >0$, and $A \sim \beta$. 

This discussion serves to put forward the following point: The characteristic timescale for which the pullback of the Wightman functions decays is $\order{\beta}$. Besides, the characteristic timescale for which the reduced density matrix of the detector varies appreciably will be clearly determined by the coupling constant $\lambda$. Actually, since the first nontrivial order in the master equations \eqref{Eq:rhodiag} is $\order{\lambda^2}$, we expect this time to be of order $\order{\lambda^{-2}}$. From these expressions, we know that $\dot{\rho}$ itself is of order $\order{\lambda^2}$, so keeping only the second order in $\lambda$ in the differential equation means that we can safely replace $\rho(\tau')$ by $\rho(\tau)$ in the integrals \eqref{Eq:rhodiag} without introducing further errors \cite{Rivas2012,Breuer2002}.  Together with a manipulation of the integrals to remove the variable $\tau$ from the integrand, this substitution transforms the integro-differential equations of motion \eqref{Eq:rhodiag} into the differential equations 
\begin{equation}\label{simple_ode}
\begin{split}
    & \Dot{\rho}_{\text{ee}}(\tau) =  - \lambda^2 I_{+} (\tau)   \rho_{\text{ee}} (\tau) + \lambda^2 I_{-} (\tau) \rho_{\text{gg}}(\tau), \\
    & \Dot{\rho}_{\text{gg}}(\tau) =  - \lambda^2 I_{-} (\tau) \rho_{\text{gg}}(\tau) + \lambda^2 I_{+} (\tau) \rho_{\text{ee}}(\tau),
\end{split}
\end{equation}
which are now local in the components of the reduced density matrix. The functions $I_{\pm}(\tau)$ can be easily seen to be
\begin{equation}\label{int}
    I_{\pm}(\tau) =  \int_{-\tau} ^{\tau} du e^{ \pm i \omega u} \mathcal{W}^{(2n)}_{D} (u).
\end{equation}
This substitution has lead to the  so-called Redfield equation \cite{Redfield1957} for each of the diagonal components of the density matrix, which is a local differential equation. However, we do not yet have a Markovian master equation in the strict sense of being a differential equation whose dynamics is described by a dynamical semigroup \cite{Rivas2012,Breuer2002}. This is because it still depends on a preferred choice of time at which we imposed initial conditions for the detector and the field, $\tau = 0 $. 

We can now make one last approximation to transform equations \eqref{simple_ode} into a Markovian equation by focusing in the physics at large $\tau$. Noticing that, in general, the integrand in \eqref{int} decays very fast at infinity [as the fall off properties of the Wightman function \eqref{falloff} ensure], we can take the integration limits to infinity in Eq. \eqref{int} without introducing a substantial error at long times. In the limit $\tau \rightarrow \infty$, the integrals \eqref{int} become the response functions with negative and positive frequency, respectively.  Therefore, we can substitute the $I_\pm (\tau)$ integrals in Eq. \eqref{simple_ode} by the associated response functions, obtaining the following differential equations with constant coefficients:
\begin{equation}
\begin{split}
    & \Dot{\rho}_{\text{ee}}(\tau) =  - \lambda^2 \mathcal{F}^{(n)}_{D} (\omega)    \rho_{\text{ee}} (\tau) + \lambda^2  \mathcal{F}^{(n)}_{D} ( - \omega) \rho_{\text{gg}}(\tau), \\
    & \Dot{\rho}_{\text{gg}}(\tau) =  - \lambda^2  \mathcal{F}^{(n)}_{D} (-\omega) \rho_{\text{gg}}(\tau) + \lambda^2  \mathcal{F}^{(n)}_{D} (\omega) \rho_{\text{ee}}(\tau).
\end{split}
\label{simple_ode2}
\end{equation}

Solutions to the differential equations \eqref{simple_ode2} exhibit a fixed point at infinite $\tau$ to which all physical solutions flow. This is a robust feature of the system of equations which persists even if we leave the $I_{\pm}$ integrals from Eq.~\eqref{simple_ode} intact. Actually, Eq. \eqref{simple_ode} can also be exactly solved, although the solution to \eqref{simple_ode2} is much more simple and will better serve the upcoming discussion. Indeed, qualitative differences between the solutions to Eq. \eqref{simple_ode} and Eq. \eqref{simple_ode2} are restricted to small deviations among the integral curves of the equations at finite times, whereas at late times solutions converge asymptotically towards the same fixed point. 

In order to solve \eqref{simple_ode2}, it is useful to introduce the function $\mathcal{H}_D^{(n)} (\omega)$ as
\begin{equation}\label{eq:jadamar}
\mathcal{H}_D^{(n)} (\omega) = \mathcal{F}_D^{(n)} (\omega) + \mathcal{F}_D^{(n)} (-\omega),
\end{equation}
in terms of which the general solution to the equations~\eqref{simple_ode2} reads
\begin{equation}
\begin{split}
     & \rho_{\text{ee}}(\tau) = 
      \frac{\mathcal{F}^{(n)}_{D} (-\omega)}{\mathcal{H}_D^{(n)} (\omega)}   \\
    & + \left( \rho_{\text{ee}} (0) - \frac{\mathcal{F}^{(n)}_{D} (-\omega)}{\mathcal{H}_D^{(n)} (\omega)} \right) \times \exp\left(- \lambda^2 \mathcal{H}_D^{(n)} (\omega)  \tau \right),
\end{split}
\label{solution}
\end{equation}
with $\rho_{\text{gg}} (\tau) = 1 - \rho_{\text{ee}} (\tau)$. In view of \eqref{solution} we observe that time-dependent terms decay exponentially fast, leading to constant values for the components of the density matrix. These asymptotic values will reveal a thermal (Gibbs) state for the detector as we see in the following. 

The fixed point at infinity is characterized by the quotient between positive and negative response functions, since they fix the ratio between the populations of the detector
\begin{equation}
\lim_{\tau \rightarrow \infty} \frac{\rho_{\text{ee}} (\tau)}{\rho_{\text{gg}} (\tau)} =  \frac{\mathcal{F}^{(n)}_{D} (\omega) }{ \mathcal{F}^{(n)}_{D} (-\omega) }.
\label{limit_infinity}
\end{equation}
We can prove that such fixed point is reached even keeping the functions $I_{\pm} (\tau)$ from Eq. \eqref{simple_ode}. Let us illustrate this aspect in detail. We make use of the normalization condition $\rho_{\text{gg}} (\tau) + \rho_{\text{ee}} (\tau)= 1$, which holds at all times due to the trace-preserving character of the equations, to rewrite \eqref{simple_ode} as a pair of decoupled differential equations. Additionally, in virtue of the normalization condition, it is sufficient to prove that one of the components of the density matrix reaches its corresponding asymptotic thermal value. Let us particularize for the $\rho_{\text{ee}}$ component obeying equation
\begin{equation}\label{Eq:rhoI}
    \dot{\rho}_{\text{ee}} (\tau) = - \lambda^2 \left[I_{\rm S}(\tau) \rho_{\text{ee}} (\tau) + I_{-} (\tau)\right],
\end{equation}
where we have made the substitution
\begin{equation}
I_{\rm S}(\tau)=I_{+}(\tau)+I_{-}(\tau).
\end{equation}
Integrating Eq. \eqref{Eq:rhoI} we have 
\begin{equation}
\begin{split}
    \rho_{\text{ee}} (\tau) = \rho_{\text{ee}}(0) e^{- \lambda^2 \int^{\tau}_{0} du I_{\rm S}(u)} +  e^{- \lambda^2 \int^{\tau}_{0} du I_{\rm S}(u)} \\
    \times \lambda^2 \int^{\tau}_{0} du I_{-} (u) e^{\lambda^2 \int^{u}_{0} dv I_{\rm S}(v)}.
    \label{explicit_sol}
\end{split}
\end{equation}
At this point, the assumption made for the fall off of the Wightman functions in Eq. \eqref{falloff} becomes relevant: The functions $I_{\pm}(\tau)$ are bounded by an exponential
\begin{equation}
I_{\pm} (\tau) - \mathcal{F}^{(n)}_{D+1} (\pm \omega) < \alpha_{\pm} e^{- \gamma_{\pm} \tau},
\end{equation}
where $\alpha_{\pm}, \gamma_{\pm}>0$. Using this bound for the functions $I_{\pm} (\tau)$ in Eq.~\eqref{explicit_sol}, we conclude that
\begin{equation}
\lim_{\tau \rightarrow \infty} \abs{ \rho_{\text{ee}} (\tau) - \frac{\mathcal{F}^{(n)}_{D} (-\omega)}{\mathcal{H}_D^{(n)} (\omega) }} =   0,
\end{equation}
hence, we have proved that the fixed point characterized by \eqref{limit_infinity} is reached asymptotically independently of the substitution made in \eqref{simple_ode2}. This proof holds under reasonable assumptions for the Wightman functions that are verified in all the cases of interest.

In view of this, the role of the response function becomes evident from the solution \eqref{solution}. We clearly see that the stationary state is reached asymptotically via the exponential decay of the $\tau$-dependent term in \eqref{solution}. The characteristic time scale $\Gamma^{-1}$ at which the non stationary features become negligible is given by 
\begin{equation}
\Gamma = \lambda^2  \mathcal{H}^{(n)}_{D} (\omega)  .
\label{decay_constant}
\end{equation}
Thus, the role of the response functions \eqref{eq:jadamar} is to modulate how fast the detector reaches the stationary state, i.e., an equilibrium state. This rate depends on the magnitude of the coupling parameter $\lambda$, which modulates the strength of the interaction. For large values of $\lambda$ (as long as they remain within the weak coupling regime) the system thermalizes more quickly, whereas, conversely, the thermalization time can be made arbitrarily long by taking a sufficiently small $\lambda$. Notice that the infinite number of degrees of freedom that characterize the field are crucial for the system to approach a fixed point. Had the role of the field in our model been played by a system with a finite number of degrees of freedom, then the state would have never reached a fixed point. In such scenario, Poincaré pseudorecurrences are unavoidable \cite{Bocchieri1957}. Actually, a full, nonperturbative treatment of the problem would result in a qualitatively different behavior of the reduced density matrix for the detector subsystem. We expect that, although the solution would approach the fixed point, it would fluctuate around it. However, we also expect such fluctuations to average to be zero, making them an extremely tiny contribution.

Notice that, up to this stage, we have considered generic fields, couplings and masses, and have restricted ourselves just to stationary trajectories. We have not invoked yet the KMS condition obeyed by the Wightman functions. Let us now focus on the two cases we are interested in comparing: A uniformly accelerating detector in vacuum and an inertial detector which is comoving with a thermal bath (a static detector with respect to the time parameter for which the Wightman functions are periodic in imaginary time). For such trajectories, the quotient between response functions is characterized by a Boltzmann factor \eqref{boltz} and the asymptotic quotient between populations reads
\begin{equation}
\lim_{\tau \rightarrow \infty} \frac{\rho_{\text{ee}} (\tau)}{\rho_{\text{gg}} (\tau)} = e^{-\beta \omega},
\end{equation}
with $\beta$ being the KMS parameter ($\beta=2\pi/a$ for the former and $\beta=1/T$ for the latter). Thus, the asymptotic stationary state of the detector in the two cases of interest is a thermal state. This same exponentially convergent behavior of the density matrix towards the Gibbs state was found in \cite{DeBievre2006}. These results are also robust in the following sense: Even if the Wightman functions are perturbatively deformed in a Lorentz-invariance-breaking way such that they no longer obey the KMS condition, the Fourier-transformed KMS condition, which is commonly written as \eqref{limit_infinity}, is still verified perturbatively \cite{Carballo2018}. Thus, the thermalization of particle detectors is also a stable phenomenon under small deformations of the relativistic quantum field theory under consideration. 

In summary, the thermalization of the detector is robust under generally reasonable approximations, which essentially involve weak coupling between the detector and the fields, an interaction that lasts long enough to thermalize, and evolution scales well separated for the detector and the interactions. The fine details of how the system approaches the thermal state are strongly dependent on the dimension, the properties of the field (self-interactions  and mass of the field) and the coupling between the field and the detector. The universality of the asymptotic thermal state for the detector is independent of these features and it is responsible for the standard interpretation of the Unruh effect. The response function for a static detector in a thermal state for the field and the response function for an accelerated detector in vacuum only agree for the particular case of a noninteracting massless scalar field in four spacetime dimensions ($D=4$). Thus, the concrete evolution of the density matrix will be different for both cases if we change some of the characteristics of the field: the dimension in which it propagates, the mass, or the interactions. It is clear then that the thermal nature of the Unruh effect does not imply that an accelerating detector in vacuum will experience the same evolution towards equilibrium that it would experience in a thermal bath, it just means that the asymptotic state reached by the detector is thermal. We hope that this work completely settles the issue of the thermal nature of the Unruh effect, i.e., in which sense it behaves as a thermal bath and in which sense it differs from it.

\section{Analysis of particular cases}
\label{Sec:Response}
In this section we will explore particular cases in order to make explicit the differences between the Unruh effect and a thermal bath. We will restrict ourselves to free fields since explicit analytic expressions are available for their response functions. In the following, we will consider couplings with parameter $n=1$, for the sake of better illustrating the differences.

We will begin by writing the response function for a real scalar scalar field in a thermal state  at temperature $T$ (this means $\beta = 1/T$, in this case) for a static detector. The explicit expression of the response function can be written \cite{Takagi1986} as
\begin{equation}
    \mathcal{\tilde{F}}_{D}^{(1)} (\omega) = \frac{\pi}{\omega}  \frac{\tilde{g}_{D}( \omega)}{e^{\beta \omega} - 1},
    \label{thermal_response}
\end{equation}
where $g_D(\omega)$ represents the density of states in Minkowski spacetime 
\begin{equation}
\tilde{g}_D(\omega) = \frac{2^{2-D} \pi^{(1-D)/2}}{\Gamma \left[ (D-1)/2 \right]} |\omega| \left( \omega^2 - m^2 \right)^{\frac{D-3}{2}} \Theta \left( |\omega| - m \right),
\end{equation}
with $m$ denoting the mass of the field. In the above expression, $\Theta(x)$ represents the Heaviside theta step function and $\Gamma(x)$ represents Euler's gamma function. The response function \eqref{thermal_response} clearly fulfills the Fourier-transformed KMS condition \eqref{boltz}. This becomes evident after noticing that $\tilde{g}_D(\omega)$ is an even function in $\omega$ and, as such, it does not contribute in any way to the quotient between excitation and decay ratios. Consequently, it is straightforward to check that relation \eqref{boltz} is satisfied, since the only timescale contribution comes from the Planckian factor times $1/\omega$ in \eqref{thermal_response}.

Let us compare this thermal response function with the one obtained for the Unruh effect. This means, let us consider a detector following an eternal trajectory of uniform acceleration $a$, being $\beta = 2 \pi /a$ the KMS parameter in this case. Let us begin with a massless scalar field. The response function reads~\cite{Takagi1986}
\begin{equation}
    \mathcal{F}_{D}^{(1)} (\omega) = \frac{\pi}{\omega}  \frac{g_{D}( \omega)}{e^{\beta \omega} - (-1)^{D}} d_{D}( \omega/a),
    \label{unruh_massless}
\end{equation}
where the functions $d_{D} (\omega/a)$ are even polynomials in $\omega$ and $g_{D}(\omega)$ is even in $\omega$ for even $D$ and odd in $\omega$ for odd $D$. Its explicit expression is
\begin{equation}
g_{D}(\omega) = \frac{2^{2-D} \pi^{(1-D)/2}}{\Gamma \left[ (D-1)/2 \right]}  |\omega|^{D-2} \left[\textrm{sign} (\omega)\right]^{D}.
\end{equation}
Notice that, for odd spacetime dimension $D$, the response function has its profile switched to a Fermi-Dirac-like distribution, instead of a Bose-Einstein-like, as originally noticed in \cite{Takagi1984}. Our definition of the functions $d_{D} (\omega/a)$ differs from the one in~\cite{Takagi1986} due to the factor $\left[\textrm{sign} (\omega)\right]^{D}$ that we have absorbed for convenience in $g_D(\omega)$. From the odd nature of the terms multiplying the Bose-Einstein factor in even $D$, we conclude that the function \eqref{unruh_massless} produces a quotient between response functions of positive and negative frequency which is characterized by a Boltzmann factor (this discussion is completely parallel to the previous case). For odd $D$, however, the factors multiplying the Fermi-Dirac distribution factor are even, and thus, irrelevant once we take the quotient between positive and negative frequency response functions. The Fermi-Dirac factor alone produces a Boltzmann factor when plugged in \eqref{boltz} and, in consequence, Eq. \eqref{unruh_massless} verifies such property for arbitrary odd~$D$. 

As we already know, the Fermi-Dirac response function appearing in odd spacetime dimension for a bosonic field does not enter into conflict with the thermal interpretation of the Unruh effect, since it just tells us how fast the detector thermalizes. Indeed, for accelerated detectors, the response function is probing properties of the field which have an influence on the path followed towards thermalization. Notice that just in the particular case $D=4$ (where the response functions for the massless real scalar field in a thermal state for a static detector and in vacuum for an accelerated detector agree) would the detector display exactly the same evolution in both situations. Indeed, from the comparison of \eqref{thermal_response} and \eqref{unruh_massless}, it is straightforward to notice that the decay rate towards equilibrium \eqref{decay_constant} is different in both cases, except for $D=4$, since $d_{4} (\omega/a) = 1$~\cite{Takagi1986}.

If we consider the Unruh effect for the free massive scalar field, although the response function does not admit a simple analytic expression in general,  in four spacetime dimensions dimensions an analytic expression in terms of $K$-Bessel functions can be found \cite{Fredenhagen1987,Kaplanek2020}. Applying similar contour deformation arguments to those explained in Appendix B of~\cite{Kaplanek2020} in four spacetime dimensions for arbitrary dimensions, one finds that an explicit expression in terms of $K$-Bessel functions and their primitives can be given~\cite{Louko}. For even spacetime dimensions an expression in terms of $K$-Bessel functions involving binomial coefficients can be found, whereas for odd spacetime dimensions one finds a sum of antiderivatives of a $K$-Bessel~\cite{Louko}. In fact, it was subject to controversy whether the response function was independent of the mass of the field until Takagi settled the question, affirming that it did depend on the mass \cite{Takagi1986}. Actually, an asymptotic expansion for $m \gg a$ is available producing the following response function:
\begin{equation}
    \mathcal{F}_{D}^{(1)} (\omega) \approx e^{- 2 m /a} \frac{  a}{8 \pi} \left( \frac{ma}{4 \pi} \right)^{D/2 - 2} e^{-  \omega \pi / a},
    \label{massive_unruh}
\end{equation}
where the Boltzmann factor appearing in the right-hand side of \eqref{boltz} is obvious given the exponential functional form of \eqref{massive_unruh}. In this case, the differences among the evolution that the detector would follow in a thermal bath \eqref{thermal_response} and from the Unruh effect \eqref{massive_unruh} characterized by the decay rate $\Gamma$ in \eqref{decay_constant} are even more striking: Whereas in the former situation the response rate has sudden vanishing of the response ratio for frequencies $\omega<m$, the latter displays a response rate suppressed by an exponential factor of the mass. This case exemplifies a scenario where a stationary detector with $\omega<m$ cannot reach equilibrium with a thermal bath, while its accelerated counterpart is able to attain it (although it takes a very long time). We conclude with this last example the discussion of particular cases. We hope that these well-known examples discussed within our formulation of the problem help in making explicit the similarities and differences between the Unruh effect and an ordinary thermal state from the perspective of detectors.  

\section{Conclusions}
\label{Sec:Conclusions}
In this work, we have revisited a well-known result in the literature whose interpretation remained obscure: the apparent inversion of statistics for a massless scalar field with odd couplings in odd spacetime dimensions. Although it was suggested that this feature implied a breakdown of the thermal interpretation of the Unruh effect, we have exposed why the content of the functional form of response function does not characterize whether the detector thermalizes or not. It is just the ratio between the response function with positive and negative frequency (the excitation to decay ratio) that characterizes whether the system reaches thermal equilibrium. Other characteristics such as dimensionality, couplings, and self-interactions of the fields, which affect the concrete form of the response function, only reflect the particular way the detector approaches equilibrium. To prove the central statement of this work, aside from Born-Markov approximation, we have assumed a polynomial fall off condition for the Wightman functions of our theory as we take the spacetime distance between points to infinity. Exploring the limitations of this assumption could be a promising future line of research.

Although generally nonanalytic for interacting theories, the quotient between positive and negative response functions is fixed to be a Boltzmann factor for an accelerated trajectory in virtue of classical arguments like the Bisognano-Wichmann theorem \cite{Bisognano1975}. In that sense, only a continuous or periodic monitorization of the detector would allow us to distinguish the specific form of the response functions. On the other hand, an asymptotic measurement of the populations of the detector will be oblivious to these distinctive features of the field under consideration and the dimensionality of the spacetime in which the detector is moving. Furthermore, it would not be able to distinguish between the Unruh effect and a standard thermal bath. We hope this work finally settles the possible doubts that might remain concerning the thermal nature of the Unruh effect in arbitrary dimensions and specifies in which sense it behaves as a thermal bath.

\acknowledgments
We would like to thank José Polo-Gómez for very useful discussions. We thank Jorma Louko and Lissa de Souza Campos for useful comments. Financial support was provided by the Spanish Government through the projects FIS2017-86497-C2-1-P, FIS2017-86497-C2-2-P (with FEDER contribution), FIS2016-78859-P (AEI/FEDER,UE), and by the Junta de Andalucía through the project FQM219. Authors J.A., C.B. and G.G.M. acknowledge financial support from the State Agency for Research of the Spanish MCIU through the ``Center of Excellence Severo Ochoa" award to the Instituto de Astrofísica de Andalucía (SEV-2017-0709). GGM acknowledges financial support from IPARCOS.

\bibliography{bunruhf_biblio}

\end{document}